\documentclass[conference,a4paper]{APSIPA2012}
\usepackage{multirow}
\usepackage{balance}
\usepackage{color}
\usepackage[numbers,sort]{natbib}
\usepackage{graphicx}
\usepackage{diagbox}
\usepackage{amsmath}
\usepackage{bm}
\usepackage[psamsfonts]{amssymb}
\usepackage{amsxtra}
\usepackage{threeparttable}

\begin{document}

\title{A Waveform Representation Framework for High-quality Statistical Parametric Speech Synthesis}

\author{%
\authorblockN{%
Bo Fan\authorrefmark{1},
Siu Wa Lee\authorrefmark{2},
Xiaohai Tian\authorrefmark{3}\authorrefmark{4},
Lei Xie\authorrefmark{1} and
Minghui Dong\authorrefmark{2}
}
\authorblockA{%
\authorrefmark{1}
School of Computer Science, Northwestern Polytechnical University, Xi'an, China \\
\authorrefmark{2}
Human Language Technology Department, Institute for Infocomm Research, Singapore\\
\authorrefmark{3}
School of Computer Engineering, Nanyang Technological University (NTU), Singapore\\
\authorrefmark{4}
Joint NTU-UBC Research Center of Excellence in Active Living for the Elderly, NTU, Singapore\\
Email: \{bofan,lxie\}@nwpu-aslp.org, \{swylee,mhdong\}@i2r.a-star.edu.sg, xhtian@ntu.edu.sg
}
}

\maketitle
\thispagestyle{empty}

\begin{abstract}
State-of-the-art statistical parametric speech synthesis (SPSS) generally uses a vocoder to represent speech signals and parameterize them into features for subsequent modeling. Magnitude
spectrum has been a dominant feature over the years. Although perceptual studies have shown that phase spectrum is essential to the quality of synthesized speech, it is often ignored by using a minimum phase filter during synthesis and the speech quality suffers. To bypass this bottleneck in vocoded speech, this paper proposes a phase-embedded waveform representation framework and establishes a magnitude-phase joint modeling platform for high-quality SPSS. Our experiments on waveform reconstruction
show that the performance is better than that of the widely-used STRAIGHT. Furthermore, the proposed modeling and synthesis platform outperforms a leading-edge, vocoded, deep bidirectional long short-term memory recurrent neural network (DBLSTM-RNN)-based baseline system in various objective evaluation metrics conducted.
\end{abstract}

\section{Introduction}
Statistical parametric speech synthesis (SPSS) has been increasingly popular due to its compact and flexible representation of voice characteristics \cite{Yos_1999_TTS_C}.
 Conventionally, in an SPSS system,
 we firstly extract parametric representations of speech including spectral and excitation parameters from a speech database and then model them with a set of models \cite{Zen_2009_SPSS_J}.
 Several statistical generative models have been applied to SPSS successfully,
 e.g., hidden Markov model (HMM)-based SPSS \cite{Zen_2009_SPSS_J}, deep neural network (DNN)-based SPSS \cite{Zen_2013_TTS_C} and deep bidirectional long short-term memory recurrent neural network (DBLSTM-RNN)-based SPSS \cite{Fan_2014_TTS_C}.

To parameterize speech signals into features for subsequent synthesis processes, vocoder has been typically used.
 It is based on the source-filter model \cite{Fant_1960_SourceFilter_B},
 which assumes a stationary speech segment is generated by passing a sound source through a vocal tract filter.
 By using a vocoder, the resultant speech features are regular and suitable for modeling.
 However, in \cite{Merritt_2014_SourceFilterInvest_C},
 their subjective listening test shows clear degradation of quality in vocoded speech.
 It further indicates that the source and filter parameters have to be jointly modelled for high-quality synthesis.
 Besides, to assure interframe coherence \cite{Cre_1997_vocoder_C}, a minimum phase hypothesis \cite{Cre_1997_vocoder_C} has been used in most vocoders, which ignores the natural mixed-phase characteristics of speech signals, resulting in apparent degradation of the speech waveform quality.


More and more works have reported the importance of phase information in different speech processing applications, such as speech synthesis \cite{Tak_2002_PhaseImportance_J,Deg_2011_PhaseImportance_J},
 iterative signal reconstruction \cite{Als_2007_IterRecons_J},
 automatic speech recognition \cite{Sch_2001_PhaseASR_C,Shi_2006_PhaseASR_J},
 speech coding \cite{Pob_2003_SpeechCoding_J} and pitch extraction \cite{Nak_2003_PitchTrack_C}.
 Paliwal et al. \cite{Pal_2003_PhaseImportance_C} have investigated the relative importance of short-time magnitude and phase spectra on speech perception through human perception listening test.
 Results show that phase spectrum clearly contributes to the speech intelligibility.
 Sometimes its contribution is as much as the magnitude spectrum.
 Koutsogiannaki et al. \cite{Kou_2014_PhaseImportance_C} have proposed the phase distortion deviation feature, enabling to capture voice irregularities and highlights the importance of the phase spectrum in voice quality assessment.
 These two works indicate that phase information is important for both human perception and voice quality assessment.
 Combining phase spectrum with magnitude spectrum in  frequency domain is equivalent to the speech waveform in time-domain.
 Therefore, the phase information is focused in our speech waveform representation framework.

There are some approaches of waveform representation directly in the time domain.
 Time domain pitch-synchronous overlap-add (TD-PSOLA) \cite{Mou_1990_TDPSOLA_J} performs pitch-synchronous analysis, modification and synthesis.
 During synthesis, speech frames are summed up.
 The quality of the reconstructed waveform with typical pitch or timing modification is similar to that of the original waveform.
 Multi-band re-synthesis pitch synchronous overlap add (MBR-PSOLA) \cite{Dut_1993_MBRPSOLA_J} comments TD-PSOLA with three mismatches:
 phase mismatch, pitch mismatch, spectral envelope mismatch.
 It further suggests to solve these mismatches by re-synthesizing voiced parts of the speech database with constant phase and constant pitch.
 The artificial processing in MBR-PSOLA decreases the quality of speech and leads to buzzy sound \cite{stylianou2001removing}.
 Alternatively, there are a few recent works for SPSS directly in the time domain. 
 Tokuda et al. \cite{Tok_2015_WaveformModel_C} have proposed an approach to model cepstral coefficients to approximate the speech waveform.
 In their framework, periodic, voiced components have not been properly generated yet.
 In \cite{maia2012complex}, complex cepstrum has been used to embed phase information for hidden semi-Markov models (HSMM) speech modelling.

In this paper, we propose a phase-embedded waveform representation framework, and establish a magnitude-phase joint modeling platform for SPSS.
 This work uses glottal-synchronous overlap add approach for speech analysis and synthesis where  glottal closure instants (GCIs) are employed.
 GCIs refer to the moments of most significant excitation that occur at the level of the vocal folds during each glottal period \cite{Smits_1995_GCI_J}.
 Short-term segments are defined as any two consecutive GCI periods.
 In order to produce smooth trajectories of our features which are required in SPSS, we design a cost function with a global smoothness constraint.
 The GCI locations selected are finally determined by conducting dynamic programming over a list of probable GCI candidates.
 Consequently, these segments will be very regular with stable magnitude and matched phase spectrum.
 With this waveform representation framework, the bottleneck suffered from vocoded speech is thus bypassed.
 This framework is hence capable of delivering better quality speech over the vocoded speech.
 Then we propose an approach for magnitude-phase joint spectrum modeling.
 Full spectrum is used in this framework, which is in line with the satisfactory performance in recent deep learning-based TTS \cite{Ling_2013_ModelSpec_J}.
 To leverage on the modeling power of deep learning,
 we use DBLSTM-RNN to learn magnitude and phase spectrum simultaneously.
 Bidirectional recurrent connections can fully exploit the speech contextual information in both forward and backward directions.
 With purpose-built memory cells to store information, the long short-term memory (LSTM) architecture does better in finding and taking advantage of the long range context.
\section{TD-PSOLA}
\label{PSOLA}
Time domain pitch-synchronous overlap add (TD-PSOLA) is used for pitch and timing modification of speech signals \cite{Mou_1990_TDPSOLA_J}, \cite{Taylor_2007_text-to-speech_B}.
 It is also popular for concatenation-based TTS.
 As no source-filter decomposition or vocoding is performed,
 the quality of resultant speech after analysis and reconstruction is highly similar to the original speech.

Given an arbitrary speech waveform signal $x(n)$,
 TD-PSOLA is carried out in the time domain.
 It first decomposes $x(n)$ into a sequence of overlapping,
 pitch-synchronized segments.
 Each segment $x_s(n)$ lasts for two pitch periods,
 running from a pitch period before and another pitch period after the segment centre.
 Then a window function $h_s(n)$, such as hanning window, will be applied to each segment.
 Assuming $S$ denotes the total number of the segments, where $s = 1, 2, ..., S$,
\begin{equation}
x_s(n) = h_s(n)x(n)
\end{equation}
$h_s(n)$ is non-zero during the above two-pitch period.
 This is how $x_s(n)$ is extracted for voiced speech;
 for unvoiced speech, the segment length is set to a constant.
 Any modification in pitch or timing can then be performed on these extracted segments.
 Finally, modified segments are overlapped and added to produce the speech output \cite{Taylor_2007_text-to-speech_B}.

Although TD-PSOLA generates pitch- and timing-modified output signals with satisfactory speech quality,
 using TD-PSOLA in speech synthesis where statistical averaging,
 modeling or signal modification are common, is not sufficient.
 This is because matched attributes on phase and pitch are needed \cite{Dut_1993_MBRPSOLA_J}.
\section{Waveform Representation Framework}
In this work, a glottal-synchronous based waveform representation framework is proposed for speech modelling. Similar to TD-PSOLA,  glottal closure instants (GCIs) represent both the pitch contours and the boundaries of individual cycles of speech.
Existing GCI detection approaches generally estimate the GCI locations in a local manner, ignoring the resultant trajectories of various acoustic attributes, i.e. segment length (representing fundamental frequency ($F0$)), magnitude and phase spectrum, exhibited in the utterance. As smooth trajectories of these attributes are necessary for SPSS, we revise a state-of-the-art GCI detection approach, so as to facilitate satisfactory modelling of these attributes.


\subsection{System Overview}
\label{WRF}
\begin{figure}[tb]
\begin{center}
\includegraphics[width=90mm]{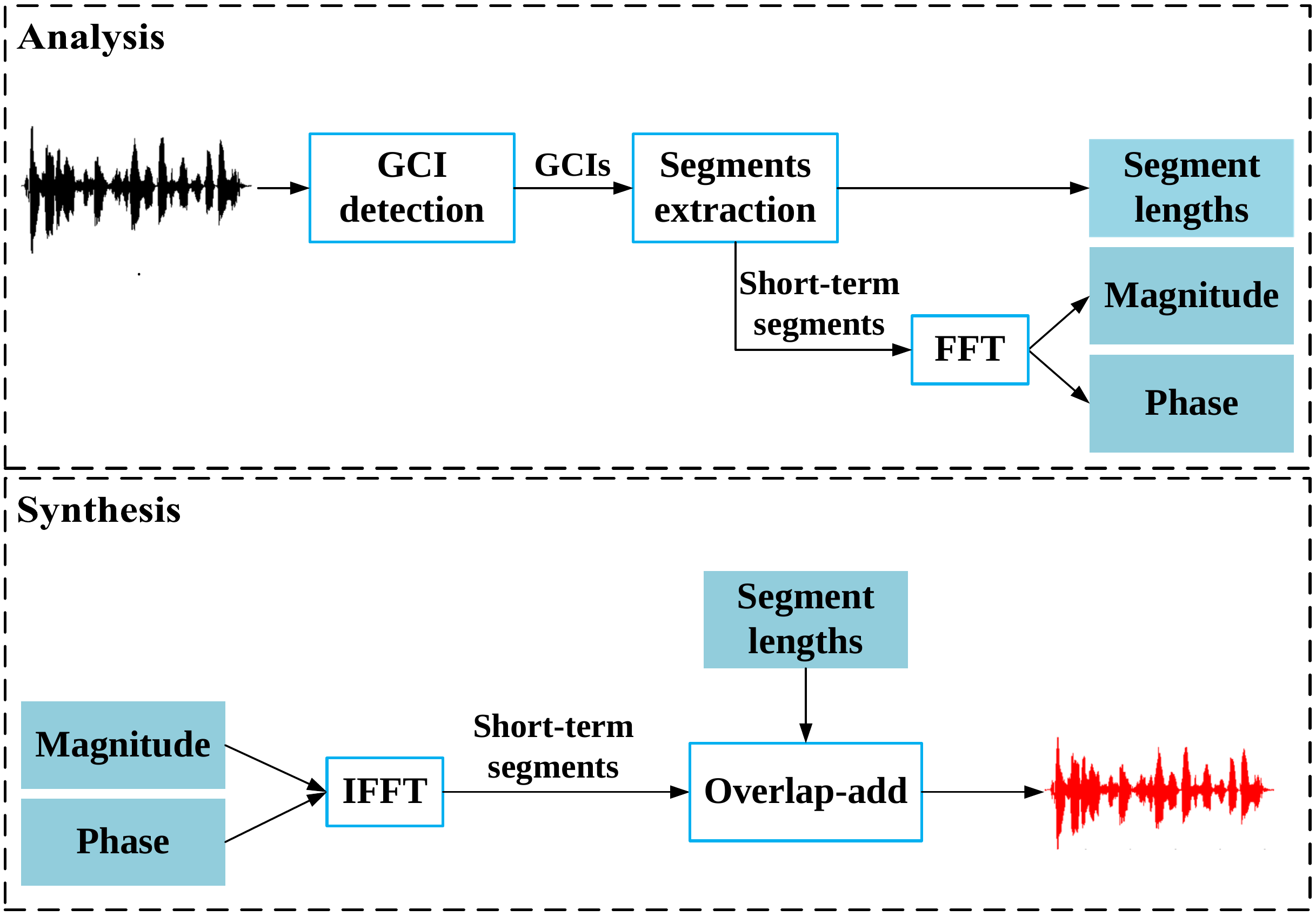}
\end{center}
\caption{Our proposed waveform representation framework.}
\label{figs:system_overview}
\end{figure}

The proposed framework, as shown in Fig.  \ref{figs:system_overview}, consists of two parts: analysis and synthesis.
In the analysis stage, given an arbitrary waveform,
firstly, the GCI locations are detected by the following revised GCI detection module.
Then, the waveform is decomposed into overlapping short-term segments. Each segment is defined by any two consecutive GCI periods.
Finally, segment lengths, magnitude and phase spectrum are used to represent these segments.

In the synthesis stage, given corresponding segment lengths, magnitude and phase spectrum,
we convert them into overlapping short-term segments.
Then, the waveform is reconstructed using the similar technique as TD-PSOLA \cite{Mou_1990_TDPSOLA_J}.

\subsection{Glottal Closure Instant Detection}

The GCI positions determine the features including segment lengths, magnitude and phase spectrum.
Thus, the GCI detection method is of great importance.

Among the present GCIs detection techniques, the Speech Event Detection using the Residual Excitation And a Mean-based Signal (SEDREAMS) algorithm \cite{Drugman_2009_SEDREAMS_C} is widely used.
In \cite{Dru_2012_GCIs_J}, SEDREAMS was shown to have the highest
robustness and reliability.
During the detection, SEDREAMS outputs only one GCI location for each GCI segment~\cite{Drugman_2009_SEDREAMS_C}. This is a local estimation process, without considering the GCI detection results in the neighborhood. However, SPSS requires smooth trajectories of speech features, which are defined once GCI locations are determined. By considering lists of probable GCI candidates and estimating the optimal GCI locations in a global manner, the trajectories of these features are stabilized.

\begin{figure}[tb]
\begin{center}
\includegraphics[width=90mm]{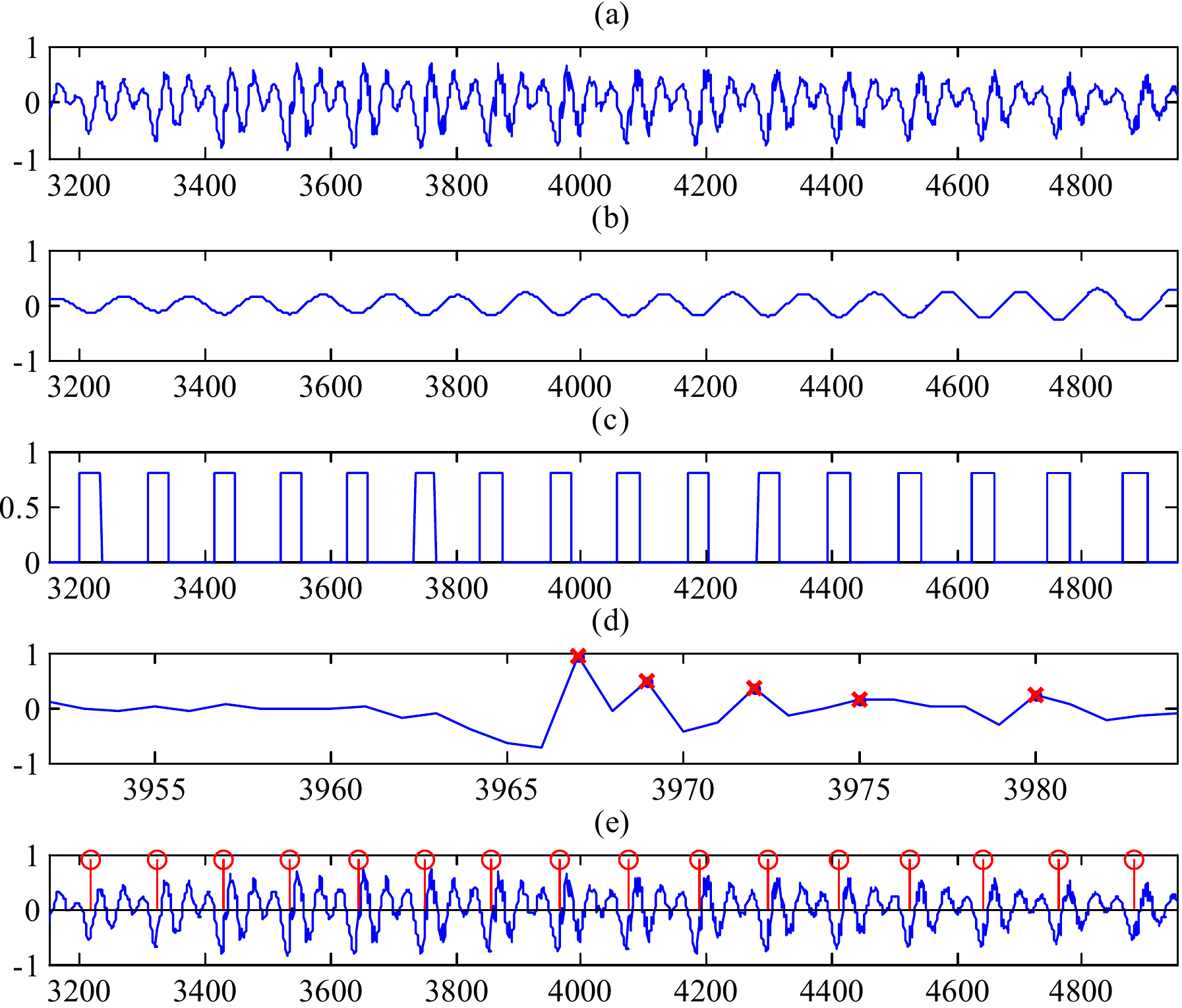}
\end{center}
\caption{(a) A section of voiced waveform; (b) The corresponding moving average signal; (c) Short intervals in the moving average signal; (d) LPC residual signal in one interval with candidates marked with red cross; (e) The final GCI locations marked with red stem.}
\label{figs:GCI_detection}
\end{figure}

Based on SEDREAMS, our modified GCI detection method contains the following steps:
\leftmargini=5mm
\begin{itemize}
	\item[a)] Given a waveform $x(n)$ (Fig. \ref{figs:GCI_detection}(a)), calculate the moving average signal (Fig. \ref{figs:GCI_detection}(b)).
	\item[b)] Determine the intervals for possible GCI locations\footnote{For the detailed implementations of the moving average filter and interval determination, please refer to~\cite{Drugman_2009_SEDREAMS_C}} (Fig. \ref{figs:GCI_detection}(c)).
	\item[c)]  $M$ candidates are chosen, based on the top $M$ highest linear predictive coding (LPC) residual values in the LPC residual signal (Fig. \ref{figs:GCI_detection}(d)), as the possible GCI locations in each interval. Suppose there are $N$ intervals, the $k$-th candidate of $i$-th interval denoted as $g_{i,k}$.
	\item[d)] Transfer all the possible segment lengths into $F0$. For the $i$-th segment, the $j$-th $F0$ is expressed as
	 \begin{equation}
 F0_{i,j} = Fs / (g_{(i+1),s} - g_{i,t}),
 \end{equation}
 where $F_s$ is the sampling frequency, $i = 1,2,...,N$, $j = 1,2,...,M \times M$, $s = 1,2,...,M$ and $t = 1,2,...,M$.
	\item[e)] Given the reference $F0^{\text{ref}}$,  the optimal segment lengths are determined by dynamic programming with the following constraint,
	 \begin{equation}
 E = {\arg} \mathop {\min }\limits_{j} \sum\nolimits_{i=1}^{N}\left\| {F0^{\text{ref}}} - {F0_{i,j}} \right\|.
 \end{equation}
 	\item[f)]  Finally, the GCI locations are deduced accordingly (Fig. \ref{figs:GCI_detection}(e)).

\end{itemize}

In our implementation, $M$ is five and the reference $F0$ is extracted by STRAIGHT~\cite{Kawahara_1999_STRAIGHT_J}.
STRAIGHT is robust for $F0$ tracking and can generate a highly accurate and smooth $F0$ trajectory.
The $F0$ trajectory extracted from STRAIGHT is robust
The dynamic programming process is implemented by the Viterbi algorithm.
In a voiced segment, the pitch located in the middle is more stable compared to the rest. Consequently, Viterbi search starts at this middle position to both ends.

\begin{figure}[htbp]
\begin{center}
\includegraphics[width=90mm]{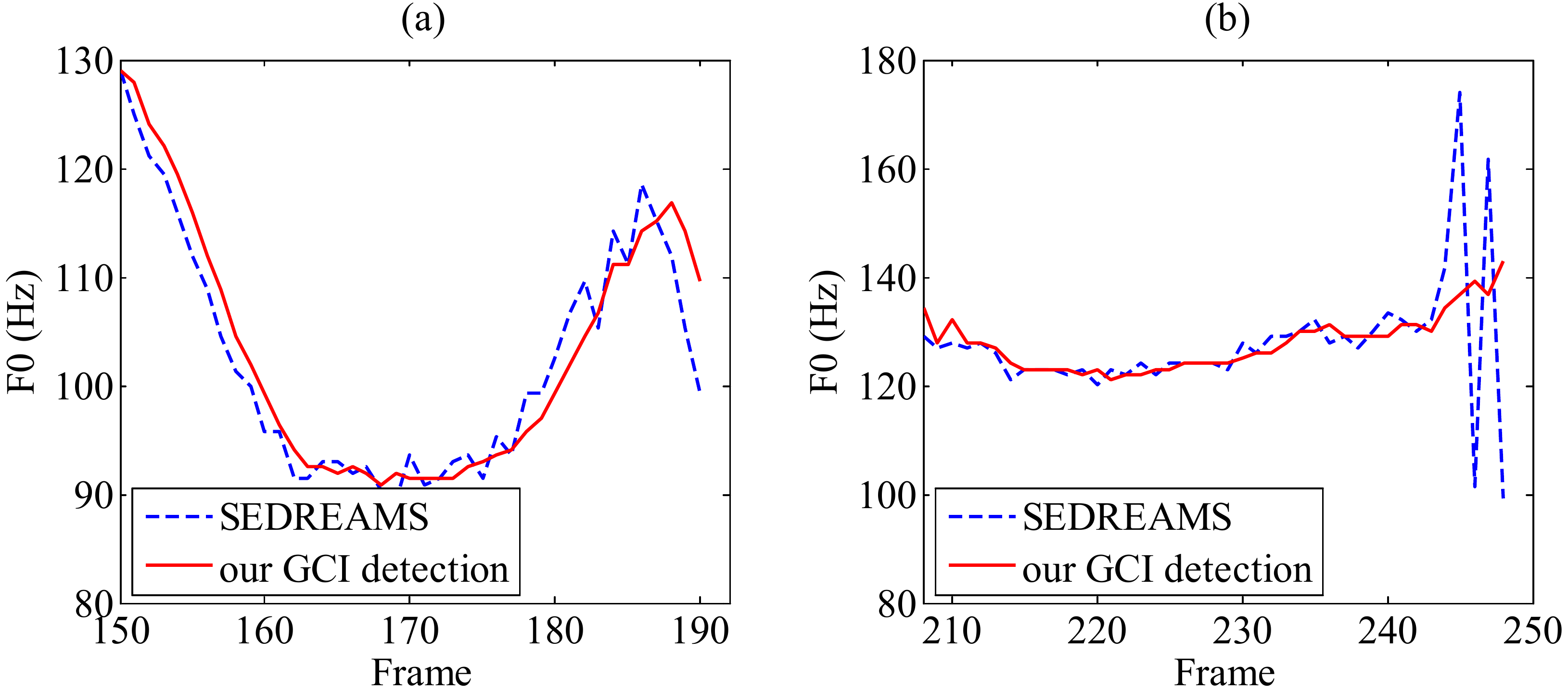}
\end{center}
\caption{A comparison of $F0$ trajectory between our GCI detection and SEDREAMS. (a) and (b) are two segments in the voiced parts.}
\label{figs:F0_trajectory}
\end{figure}

A comparison of $F0$ trajectory between our GCI detection and SEDREAMS is depicted in Fig. \ref{figs:F0_trajectory}.
 From Fig. \ref{figs:F0_trajectory}(a),
 it is observed that the $F0$ given by our GCI detection is smoother than the one from SEDREAMS.
 And from Fig. \ref{figs:F0_trajectory}(b), it is clear that our GCI detection approach removes some abnormal jumps (around the 247-th frame) of the $F0$ trajectory occurred in the SEDREAMS.

 \begin{figure*}[thbp]
\begin{center}
\includegraphics[width=180mm]{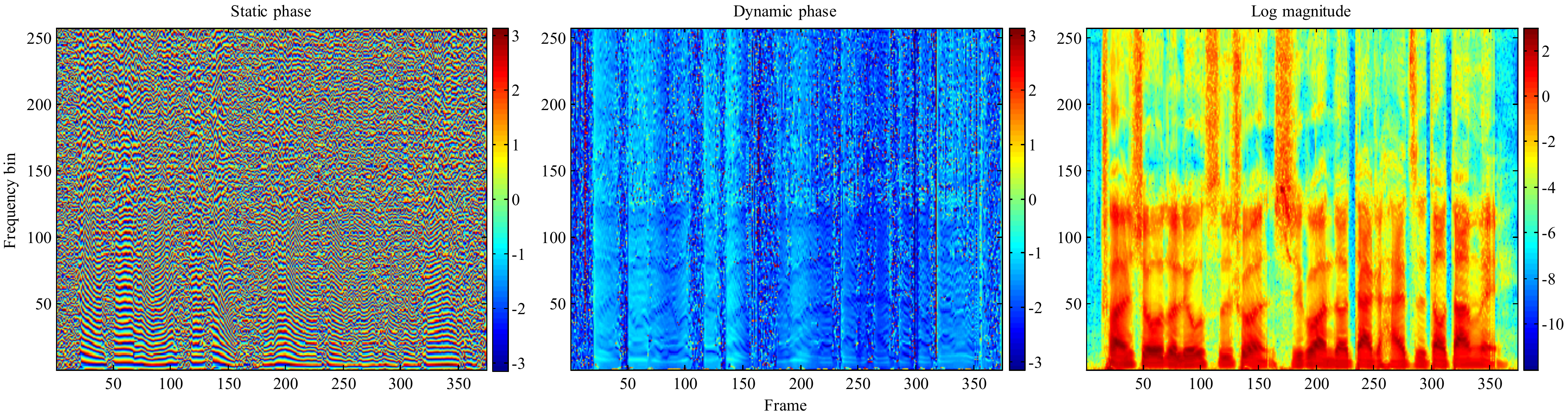}
\end{center}
\caption{Static and dynamic phase spectrum as well as their corresponding magnitude spectrum.}
\vspace*{-3pt}
\label{figs:complex}
\end{figure*}
\section{Waveform Modeling}
State-of-the-art SPSS usually models the magnitude spectrum of speech signals and discards the phase spectrum.
 During synthesis, a vocoder based on minimum-phase or zero-phase filter is often used together with the generated magnitude spectra to produce the synthesized output.
 Nevertheless, phase spectrum has been recently found to be essential for speech perception.
 The speech quality of vocoded outputs are found to be degraded from the original speech recordings \cite{Merritt_2014_SourceFilterInvest_C}.
 This may shed light on SPSS, where speech waveform with phase information in addition to the existing magnitude spectrum, is modeled.

In our work, speech signals are modeled by the corresponding magnitude and phase spectra,
 without the use of a vocoder.
 Consequently, reconstruction of speech waveform is facilitated.
 We use a recently-emerging learning technique, DBLSTM-RNN, to jointly model the two spectra.
 DBLSTM-RNN is well-suited for learning sequential events apart from long time lags of unknown size \cite{graves2012supervised}.
 Promising performance in various speech applications is observed \cite{Graves_2013_ASR_C}, \cite{Fan_2014_TTS_C}.

Our joint model of magnitude and phase is constructed as follows.
 We employ line spectrum pair (LSP) as the feature representation of magnitude spectrum.
 LSP, being an alternative LPC spectral representation, is robust and suitable for interpolation and modeling \cite{Soong_1984_LSP_compression_C}, \cite{Itakura_1975_LSP_J}.

For phase spectrum, we propose to use the dynamic phase spectrum for this waveform learning TTS framework.
 It is also called group delay:
 the group delay $\tau_k(n)$ at time $n$ and frequency bin $k$ is calculated as the frequency derivative of the instantaneous phase $\theta_k(n)$, i.e.
\begin{equation}
\tau_k(n) = \theta_{k}(n) - \theta_{k-1}(n).
\end{equation}

To enable reconstruction of the phase spectrum after DBLSTM-RNN modeling,
 the instantaneous phase at the first frequency bin is kept,
 together with the group delays of the remaining frequency bins.
 In other words, our phase representation consists of $\theta_1(n), \tau_2(n), \tau_3(n), ..., \tau_K(n)$,
 where $K$ is the total number of frequency bins.

This group-delay-based phase representation is found to be stable and facilitates statistical modeling in subsequent TTS process, as shown in Fig.~\ref{figs:complex}.
 Comparing the spectra of static phase and dynamic phase,
 the distribution of the dynamic phase often exhibits a smaller range.
 Comparing the log magnitude spectrum with the dynamic phase spectrum,
 patterns of voiced and unvoiced portions are consistent and spectral patterns of individual speech sounds are quite similar in the log magnitude spectrum and the dynamic phase spectrum.
 This is important and useful for our joint modeling.
 On the contrary, there is no clear difference in the static phase spectrum for individual speech sounds.
 When moving along the time-axis, the static phase spectra look like the same.
\section{Experiments}
\label{EXP}
We conducted two experiments to assess the efficacy of our waveform representation framework.
 In the experiment on waveform reconstruction,
 objective and subjective evaluations were carried out to compare the performance between our framework and other three vocoders: STRAIGHT, Tandem-STRAIGHT~\cite{Kawahara_2008_tandem_C} and AHOCoder~\cite{Erro_2014_ahocoder_J} respectively.
 As we know, STRAIGHT is a very popular vocoder used for speech analysis and reconstruction, and Tandem-STRAIGHT is the upgrade version of STRAIGHT.
 AHOCoder is reported to be of similar quality compared with STRAIGHT.
 In the experiment on waveform modeling,
 we trained a text-to-speech (TTS) system based on our framework and also a baseline TTS system \cite{Fan_2014_TTS_C} as a comparison.
 This baseline is a leading-edge approach based on DBLSTM-RNN and generates high-quality synthesized speech.
 It uses STRAIGHT as its vocoder.

A corpus with 4,936 Chinese utterances (around 6 hours) spoken by a native male speaker in a neutral style was used in our experiments.
 Speech waveform signals are sampled at 16kHz.
 The contextual labels are both phonetically and prosodically rich, including quin-phone, prosody, tone and syllable information.
 For TTS systems, the training, validation and test data consist of 3,949, 494 and 493 utterances, respectively.
\subsection{Experiment on Waveform Reconstruction}
\label{exp:waveRe}
Speech waveform in the test set of the corpus was analyzed and re-synthesized using our waveform representation framework and the three vocoders.
 The reconstructed speech waveform was then used for objective and subjective evaluations.
\subsubsection{Objective Evaluation}
In the objective evaluation, we calculated the root mean square error (RMSE) between the reconstructed and original speech waveform signals in the voiced parts (RMSE\_voiced), the unvoiced parts (RMSE\_unvoiced) and the entire waveform (RMSE), respectively.
 The results are shown in Table \ref{table:exp_recon}.
 These voiced/unvoiced results from our framework and the three vocoders generally represent the performance on vowels/consonants respectively.
\begin{table} [ht]
\caption{\label{table:exp_recon} {Reconstruction performance: our framework vs. the three vocoders}}.
\centerline{
\begin{tabular}{|c|c|c|c|}
\hline
\diagbox[width=12em]{Methods}{Measures} & RMSE\_voiced & RMSE\_unvoiced & RMSE \\
\hline
Our framework & \bf{0.026} & \bf{0.042} & \bf{0.031} \\
\hline
STRAIGHT~\cite{Kawahara_1999_STRAIGHT_J} & 0.173 & 0.044 & 0.152 \\
\hline
Tandem-STRAIGHT~\cite{Kawahara_2008_tandem_C} & 0.177 & 0.044 & 0.156\\
\hline
AHOCoder~\cite{Erro_2014_ahocoder_J} & 0.182 & 0.049 & 0.160\\
\hline
\end{tabular}}
\end{table}
\begin{figure}[htbp]
\begin{center}
\includegraphics[width=90mm]{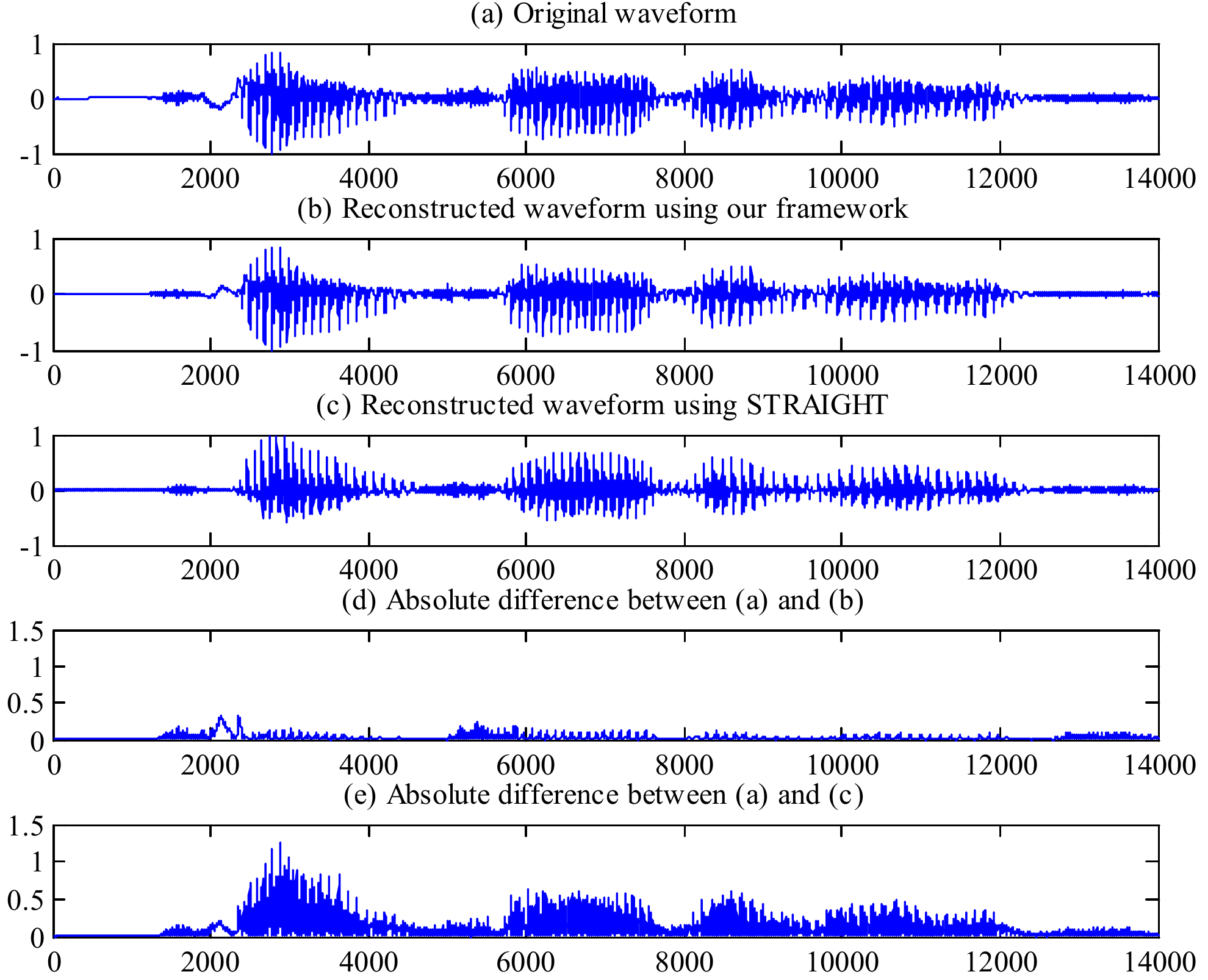}
\end{center}
\caption{The reconstructed waveform using our framework and STRAIGHT.}
\label{figs:exp_recon}
\end{figure}

The objective evaluation result shows that the performance of our framework is much better than that of the three vocoders especially in the voiced parts.
 The short-term segments are extracted at a constant rate in the unvoiced parts from our framework which is similar to STRAIGHT.
 Taking the waveform in Fig. \ref{figs:exp_recon} around the 5000-th sample as an example,
 the absolute difference between (a) and (b) is very close to that between (a) and (c).
 In the voiced parts, our framework performs much better than STRAIGHT does.
 It is because our framework retains the full phase spectrum, while STRAIGHT discards it and uses a minimum-phase setting instead.
 We can see clearly from Fig. \ref{figs:exp_recon} that the absolute difference between (a) and (b) is much smaller than that between (a) and (c) in the voiced parts.
\subsubsection{Subjective Evaluation}
\begin{figure}[htbp]
\begin{center}
\includegraphics[width=80mm]{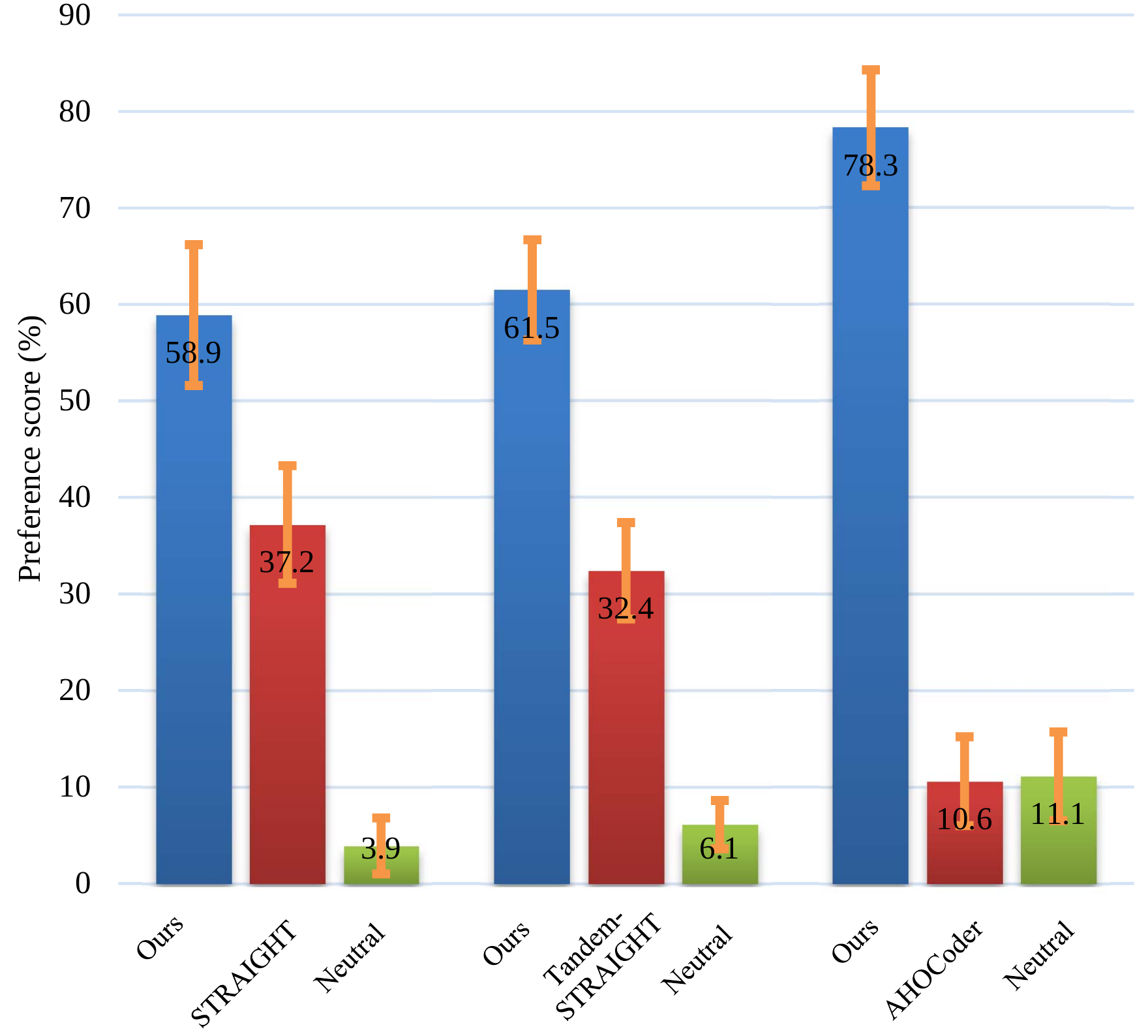}
\end{center}
\caption{The ABX result of the reconstructed speech waveform using our framework and the three vocoders. We conducted $t$-test using a significance level of $p < 0.05$ which is depicted with the error bars in orange.}
\label{figs:ABX1}
\end{figure}

20 pairs of speech waveform are randomly selected from the reconstructed waveforms.
 Then a group of 20 subjects were asked to perform the ABX preference test.
 We put the original waveform into {\bf X},
 while we put the waveform reconstructed using our framework and each of the three vocoders into {\bf A} and {\bf B} randomly.
 Each subject was asked to answer which one({\bf A} or {\bf B}) is more similar to {\bf X}.
 The third option {\bf {Neutral}} means the subject has no preference on A or B.
 The ABX result is shown in Fig. \ref{figs:ABX1}.
 We can clearly see that the reconstructed speech waveform using our framework is significantly preferred as compared with all of the three vocoders.
\subsection{Experiment on Waveform Modeling}
In the baseline DBLSTM-RNN-based TTS \cite{Fan_2014_TTS_C},
 STRAIGHT is used to vocode the speech waveform by a 25-ms moving window, and shifted every 5-ms.
 The generated magnitude spectrum from STRAIGHT was converted into LSP.
 The dimensionality of the input contextual label is 427.
 The output feature contains voiced/unvoiced flag (1 dimension), log F0 (1 dimension), LSP (40 dimensions) and gain (1 dimension), totally 43 dimensions.
 As suggested in \cite{Fan_2014_TTS_C}, a neural network with two BLSTM layers sitting on two feed forward layers with 256 nodes in each layer is employed to train the DBLSTM-RNN-based TTS.

For our TTS system,
 features were extracted from the short-term segments specified by GCI locations.
 The format of the input label is the same as the baseline.
 The segment length is transformed into $F0$.
 The output feature comprises several components: voice/unvoiced flag (1 dimension), log $F0$ (1 dimension), LSP (40 dimensions), gain (1 dimension) and dynamic phase feature (257 dimensions), totally 300 dimensions.
 The same network topology as baseline is used to train our TTS system.

To evaluate the performance of these two TTS systems, five metrics are used for objective evaluation:
\begin{itemize}
\item RMSE\_$F0$: root mean square error in $F0$ estimation;
\item Voiced/unvoiced (V/U) error rate;
\item Log spectral distance (LSD):
\begin{multline}
LSD({\bf S}_p,{\bf S}_g)=\\
\sqrt {\frac{1}{N}\sum_{j=1}^{N}(\sum_{k=1}^{M_{s}}[10log_{10} \, s_p(j,k)-10log_{10} \, s_g(j,k)]^2)},
\end{multline}
where ${\bf S}_p$ and ${\bf S}_g$ are the predicted and ground-truth magnitude spectrum, respectively.
 $N$ is the total number of frames in the voiced parts and $M_s$ refers to the dimensionality of magnitude spectrum.
 $s_p(j,k)$ is the the $k$-th value of magnitude in $j$-th frame;
\item Mel cepstral distance (MCD):
\begin{equation}
MCD({\bf c}_p,{\bf c}_g)=\frac{10}{ln \, 10}{\sqrt {2\sum_{k=1}^{M_c}[c_p(k)-c_g(k)]^2}},
\end{equation}
where ${\bf c}_p$ and ${\bf c}_g$ are the predicted and ground-truth Mel cepstrum coefficient vectors, respectively, and $M_c$ refers to the dimensionality of Mel cepstrum coefficients;
\item Dynamic phase distance (DPD):
\begin{equation}
DPD({\bf d}_p,{\bf d}_g)=\sqrt {\sum_{k=1}^{M_{d}}[d_p(k)-d_g(k)]^2},
\end{equation}
where ${\bf d}_p$ and ${\bf d}_g$ are the predicted and ground-truth dynamic phase feature vectors, respectively, and $M_{d}$ refers to the dimensionality of the dynamic phase feature.
\end{itemize}

The synthesized speech waveform from the labels in the test set uses the ground-truth durations.
 These five metrics are calculated at the GCIs level, i.e., the short-term segments are specified by the GCIs locations.
 In order to make the systems comparable,
 GCI detection is required for all speech waveforms synthesized from any system under comparison.
 And after the GCI detection, it should be aligned to the ground-truth GCIs by finding out the closest one.

The objective evaluation result is shown in Table \ref{table:exp_modeling}.
 It shows that our TTS system is better than the baseline in terms of all the five metrics.
 In particular, for DPD, the average absolute difference in one frequency bin is about 0.70rad in our TTS system while 0.91rad for the baseline TTS system.
 \begin{table} [ht]
\caption{\label{table:exp_modeling} {Objective evaluation on waveform modeling with $t$-test using a significance level of $p < 0.05$.}}
\centerline{
\begin{tabular}{|c|c|c|}
\hline
\diagbox[dir=SE]{Measures}{Methods} & Our TTS system & Baseline~\cite{Fan_2014_TTS_C} \\
\hline
RMSE\_F0 (Hz) & \bf{23.6 $\pm$ 1.2} & 28.0 $\pm$ 2.3\\
\hline
V/U error rate (\%) & \bf{5.9 $\pm$ 0.6} & 8.6 $\pm$ 1.0\\
\hline
LSD (dB) & \bf{59.2 $\pm$ 0.8} & 63.9 $\pm$ 1.1 \\
\hline
MCD (dB) &  \bf{4.5 $\pm$ 0.1} & 4.8 $\pm$ 0.1 \\
\hline
DPD (rad) & \bf{11.4 $\pm$ 0.2} & 14.7 $\pm$ 0.3\\
\hline
\end{tabular}}
\end{table}

\section{Conclusions and Future Work}
This paper proposed a glottal-synchronous based waveform representation framework for high-quality statistical parametric speech synthesis. Speech signal was represented by magnitude and phase full-spectral components, without the the use of a vocoder. We revised the SEDREAMS GCI detection approach to improve the feature stability for statistical modelling.
Both objective and subjective evaluations were conducted to assess the reconstruction performance of our framework.
Results indicate that, comparing to the reconstructed signal obtained by three popular vocoders, the proposed framework achieves promising results in RMSE in time domain speech waveform and preference score.

We also proposed a platform for speech modelling. DBLSTM-RNN is applied to jointly model the corresponding magnitude and phase spectra, and group delay-based phase representation is used to facilitate statistical modelling.
Objective results show that, the TTS system based on the proposed framework generates the features, specifically the phase feature, with lower distortion as compared with a vocoder based system.
Further works include studying the speech quality of synthesized speech and the associated factors and experiments on subjective evaluation.
\section*{Acknowledgment}
This work was supported by the National Natural Science Foundation of China (61175018 and 61571363).

\bibliographystyle{IEEEbib}
\balance
\bibliography{ref}

\end{document}